\documentclass[aps,prd,twocolumn,superscriptaddress]{revtex4}
\usepackage{epsfig,epsf}
\usepackage{amsmath}
\usepackage{amsthm}
\usepackage{amsfonts}
\usepackage{amssymb}
\usepackage{dsfont}
\usepackage{multirow}
\usepackage{appendix}
\usepackage{slashed}
\usepackage[active]{srcltx}
\usepackage{psfrag}

\begin{document}

\title{Analysis of the structure of  $\Xi(1690)$ through its decays }
\date{\today}
\author{T.~M.~Aliev}
\affiliation{ Physics Department, Middle East Technical
University, 06531 Ankara, Turkey}
\author{K.~Azizi}
\affiliation{School of Physics, Institute for Research in
Fundamental Sciences (IPM), P. O. Box 19395-5531, Tehran, Iran}
\affiliation{Department of Physics, Do\v{g}u\c{s} University,
Acibadem-Kadik\"{o}y, 34722 Istanbul, Turkey}
\author{H.~Sundu}
\affiliation{Department of Physics, Kocaeli University, 41380 Izmit, Turkey}

\begin{abstract}
The mass and pole residue of the  first orbitally and radially
excited $ \Xi$ state as well as the ground state residue are
calculated by means of the two-point QCD sum rules. Using the obtained
results for the spectroscopic parameters,  the strong coupling constants relevant to the decays
$\Xi(1690)\rightarrow \Sigma K$ and $\Xi(1690) \rightarrow \Lambda
K$ are calculated within the light-cone QCD sum rules and width of
these decay channels are estimated. The obtained results for the mass
of $ \widetilde{\Xi} $ and ratio of the $Br\Big(\widetilde{\Xi}\rightarrow \Sigma
K\Big)/Br\Big(\widetilde{\Xi}\rightarrow \Lambda K\Big)$, with $ \widetilde{\Xi} $ representing the orbitally excited state in $ \Xi $ channel,  are in
nice agreement with the experimental data of the Belle
Collaboration. This allows us to conclude that the $\Xi(1690)$ state,
most probably, has negative parity.
\end{abstract}

\maketitle

\section{Introduction}

Understanding the spectrum of baryons and looking for new baryonic states
constitute one of the main research directions in hadron physics.
Impressive developments of experimental techniques allow 
discovery of many new hadrons. Despite these developments, the spectrum
of $\Xi$ baryon is still not well established. This is due to the
absence of high intensity anti-kaon beams and small production rate
of the $\Xi$ resonances. At present time only the ground state
octet and decuplet baryons as well as $\Xi(1320)$ and $\Xi(1530)$
baryons are well established. Up to present time the quantum
numbers of $\Xi(1690)$, $\Xi(1820)$ and $\Xi(1950)$ have not been
determined. Theoretically, the spectrum of $\Xi$ baryon, within
different approaches, have been studied intensively (see
\cite{Isgur:1978xj,Oh:2007cr,Lee:2002jb,Pervin:2007wa,Melde:2008yr,Schat:2001xr,Xiao:2013xi,Chao:1980em,Capstick:1986bm}
and references therein).

The main results of these studies are that  different
phenomenological models explain successfully the nature of
$\Xi(1320)$ and $\Xi(1530)$ states. However, these approaches predict
controversially results for other excitations of $\Xi$ baryons. In
\cite{Chao:1980em} using the nonrelativistic quark model the mass
of $\Xi(1690)$ is calculated and it is obtained that it might be
radial excitation of $\Xi$ with $J^P=\frac{1}{2}^+$. This result
was then supported by the quark model calculations in
\cite{Melde:2008yr}. However within the relativistic quark model
in \cite{Capstick:1986bm} it was established that the first radial
excitation should have mass around $1840~\mathrm{MeV}$. In
\cite{Pervin:2007wa} the authors suggested that the $\Xi(1690)$
state might be orbital excitation of $\Xi$ with
$J^P=\frac{1}{2}^-$. This point of view was supported by
calculations performed within Skyrme model \cite{Oh:2007cr} and
chiral quark model \cite{Xiao:2013xi}.
The controversy results suggests independent analysis for
establishing the nature of $\Xi(1690)$ state.

In the present study, within the light cone QCD sum rules, we estimate the
widths of the $\Xi\rightarrow \Lambda K$ and $\Xi\rightarrow \Sigma K$
transitions. We suggest that $\Xi(1690)$ state may be radial ($ \widetilde{\Xi} $) or
orbital ($\Xi'$) excitation of $\Xi$ baryon. For establishing these decays
 we need to know the residue of $\Xi(1690)$ as well as the
strong coupling constants for these decays. For calculation of the
mass and residue of the $\Xi$ states as the main inputs of the calculations we employ the two point QCD sum rule method.

The paper is arranged as follows. In Section II the mass and
residue of $\Xi(1690)$ baryon within both scenarios, namely considering $\Xi(1690)$
as the orbital and radial excitations of $\Xi$ baryon, are
calculated. In section III  we present the calculations of the strong
coupling constants defining the  $\Xi(1690)\rightarrow \Sigma (\Lambda) K$
transitions within both scenarios. By using  the obtained
results for the coupling constants we estimate the relevant decay widths and
compare our predictions on decay widths with the existing experimental
data in this section, as well. We reserve section IV for the concluding remarks and some lengthy expressions are moved to the Appendix.


\section{Mass and Pole Residue of the  first orbitally and radially excited $\Xi$ state}

For calculation of the widths of $\Xi\rightarrow \Sigma K$ and
$\Xi\rightarrow \Lambda K$ decays we need to know the residues of
$\Xi$, $\Sigma$ and $\Lambda$ baryons. In present work we consider
two possible scenarios about nature of the $\Xi(1690)$: a) it is represented
as radial excitation of the ground state $\Xi$. In other words it carries
the same quantum numbers as the ground state $\Xi$, i.e. $J^P=\frac{1}{2}^+$. b)
The $\Xi(1690)$ state is considered as first orbital excitation of the ground state $\Xi$, i.e.
it is negative parity baryon with $J^P=\frac{1}{2}^-$. In the following we will try to answer the question that which
scenario is realized in nature? To answer this question we will
calculate the mass of $\Xi(1690)$ state and decay width of the $\Xi \rightarrow \Sigma K$ and $\Xi
\rightarrow \Lambda K$ transitions and then compare the ratio of these decays as well as the prediction on the mass
with existing experimental data. Note that the BABAR Collaboration
has measured the mass ($ m=1684.7\pm1.3^{+2.2}_{-1.6} $) MeV and width ($ \Gamma=8.1 ^{+3.9+1.0}_{-3.5-0.9}$) MeV of
 $\Xi(1690)$ \cite{Aubert:2008ty}
and Belle Collaboration has measured the mass ($ m=1688\pm 2$) MeV and width ($ \Gamma=11\pm 4$) MeV  of this state as well as  the ratio
$\frac{B\Big(\Xi(1690)^0\rightarrow K^-\Sigma^+
\Big)}{B\Big(\Xi(1690)^0\rightarrow \bar{K}^0 \Lambda^0 \Big)}$.
The experimental value for this ratio  measured by Belle is $0.50\pm0.26$
\cite{Abe:2001mb}.

For determination of the mass and residue of  $\Xi$ baryon, we start
with the following two point correlation function:
\begin{equation}
\Pi (q)=i\int d^{4}xe^{iq\cdot x}\langle 0|\mathcal{T}\Big\{\eta_{\Xi }(x)%
\bar{\eta}_{\Xi}(0)\Big\}|0\rangle ,  \label{eq:CorrF1}
\end{equation}%
where $\eta_{\Xi}(x)$ is the interpolating current for $\Xi $
state with spin $J^P=\frac{1}{2}^+$ and $\mathcal{T}$ indicates
the time ordering operator. The general form of the interpolating
current for the  spin-${1\over 2}$ $\Xi$ baryon can be
written as \cite{Chung,Dosch}:

\begin{eqnarray}
 \label{Eq:Current}
\eta_{\Xi} &=&\epsilon^{abc}
\Big\{\Big(s^{T,a}(x)Cu^b(x)\Big)\gamma_5s^c(x) \nonumber \\
&+&\beta\Big(s^{T,a}(x)C\gamma_5u^b(x)\Big)s^c(x) \Big\}~,
\end{eqnarray}
 where $a,b,c$
are the color indices and  $\beta$ is an arbitrary parameter with
$\beta =-1$ corresponding to the Ioffe current. $C$ is the charge
conjugation operator.

According to the general philosophy of QCD sum rules method, for calculation of the mass
and residue of $\Xi$ baryons the correlation function needs to
be calculated in two different ways: a) in terms of hadronic degrees of
freedom and b) in terms of perturbative and
vacuum-condensates contributions expressed as functions of QCD degrees of freedom in deep- Euclidian domain $q^2\ll
0$. After equating these two representations,  the desired QCD sum rules
for the physical quantities of the baryons under consideration are
obtained. As already noted, the quantum numbers $J^P$ of
$\Xi(1690)$ state have not been  determined via experiments yet. Therefore,
firstly we consider the case when $\Xi(1690)$ represents  a negative
parity baryon. The hadronic side of the correlation function is
obtained by inserting  complete sets of relevant intermediate states. For
calculation of the hadronic side of the correlation function, we
would like to note that the above interpolating current has nonzero
matrix element with baryons of both parities. Taking into account
this fact and saturating the correlation function by complete sets of
intermediate states with both parities we obtain:
\begin{eqnarray}
\Pi ^{\mathrm{Phys}}(q) &=&\frac{\langle 0|\eta|\Xi (q,s)\rangle
\langle
\Xi (q,s)|\overline{\eta}|0\rangle }{m^{2}-q^{2}}  \notag \\
&+&\frac{\langle 0|\eta|\widetilde{\Xi }(q,\widetilde{s})\rangle
\langle
\widetilde{\Xi }(q,\widetilde{s})|\overline{\eta}||0\rangle }{\widetilde{m}%
^{2}-q^{2}}  \notag \\
&+&\ldots ,  \label{eq:CF1/2}
\end{eqnarray}%
where $m$, $\widetilde{m}$ and $s$, $\widetilde{s}$ are the masses
and spins of the ground and first orbitally excited $\Xi $
baryons, respectively. Here dots represent the contributions of
higher states and continuum.

The matrix elements in Eq. (\ref{eq:CF1/2}) are determined as
\begin{eqnarray}
\langle 0|\eta|\Xi (q,s)\rangle  &=&\lambda u(q,s),  \notag \\
\langle 0|\eta|\widetilde{\Xi }(q,\widetilde{s})\rangle  &=&\widetilde{%
\lambda }\gamma _{5}u(q,\widetilde{s}).  \label{eq:MElem}
\end{eqnarray}%
Here $\lambda $ and $\widetilde{\lambda }$ are the residues of the
ground and first orbitally excited $\Xi $ baryons, respectively.
Using Eqs.\ (\ref {eq:CF1/2}) and (\ref{eq:MElem}) and performing
 summation over the spins of corresponding baryons, we obtain
\begin{equation*}
\Pi ^{\mathrm{Phys}}(q)=\frac{\lambda^2 (\slashed q+m)}{m^{2}-q^{2}}+\frac{%
\widetilde{\lambda }^2(\slashed q-\widetilde{m})}{\widetilde{m}^{2}-q^{2}}%
+\ldots.
\end{equation*}
We perform  Borel transformation in order to suppress the
contribution of higher state and continuum,
\begin{eqnarray}
\mathcal{B}\Pi ^{\mathrm{Phys}}(q) &=&\lambda ^{2}e^{-\frac{m^{2}}{M^{2}}}(%
\slashed q+m)  \notag \\
&+&\widetilde{\lambda }^2e^{-\frac{\widetilde{m}^{2}}{M^{2}}}(\slashed q-%
\widetilde{m}).  \label{eq:Bor1}
\end{eqnarray}
where $M^2$ is the square of Borel mass parameter.

The correlation function from QCD side can be calculated by
inserting Eq. (\ref{Eq:Current}) to Eq. (\ref{eq:CorrF1}) and usage of
Wick's theorem to contract the quark fields. As a  result we have an expression in terms of the involved quark propagators having
 perturbative and non-perturbative
contributions. For calculation of these contributions we need explicit
expressions of the light quark propagators.  By using the light quark
propagators in the coordinate space and performing the Fourier and
Borel transformations, as well as performing the continuum
subtraction by using the hadron-quark duality ansatz, after  lengthy
calculations, for the correlation function we obtain
\begin{equation}
\mathcal{B}\Pi ^{\mathrm{QCD}}(q)=\mathcal{B}\Pi
_{1}^{\mathrm{QCD}} \slashed q+\mathcal{B}\Pi
_{2}^{\mathrm{QCD}}I,
\end{equation}
where, the  expressions  for  $\mathcal{B}\Pi _{1}^{\mathrm{QCD}}$ and
$\mathcal{B}\Pi _{2}^{\mathrm{QCD}}$ are presented in Appendix.

Having calculated both the hadronic and QCD sides of the
correlation function, we match the coefficients of the
corresponding structures $\slashed q$ and $I$ from these
representations to find the following sum rules:
\begin{eqnarray}
\lambda^{2}e^{-\frac{m^{2}}{M^{2}}}+\widetilde{\lambda}^{2}e^{-\frac{
\widetilde{m}^{2}}{M^{2}}}&=&\mathcal{B}\Pi _{1}^{\mathrm{QCD}},
\notag
\\
\lambda
^{2}me^{-\frac{m^{2}}{M^{2}}}-\widetilde{\lambda}^{2}\widetilde{m}e^{-\frac{
\widetilde{m}^{2}}{M^{2}}}&=&\mathcal{B}\Pi _{2}^{\mathrm{QCD}}.
\label{eq:MFor1}
\end{eqnarray}
From these equations one can easily find:
\begin{eqnarray}
\widetilde{m}^{2}&=&\frac{\mathcal{B}\widetilde{\Pi}
_{2}^{\mathrm{QCD}}-m\mathcal{B}\widetilde{\Pi}
_{1}^{\mathrm{QCD}}}{\mathcal{B}\Pi
_{2}^{\mathrm{QCD}}-m\mathcal{B}\Pi _{1}^{\mathrm{QCD}}},
\nonumber \\
\widetilde{\lambda}^{2}&=&\frac{m\mathcal{B}\Pi
_{1}^{\mathrm{QCD}}-\mathcal{B}\Pi
_{2}^{\mathrm{QCD}}}{m+\widetilde{m}}e^{\frac{\widetilde{m}^2}{M^2}},
\nonumber \\
\lambda^2&=&\frac{\widetilde{m}\mathcal{B}\Pi
_{1}^{\mathrm{QCD}}+\mathcal{B}\Pi
_{2}^{\mathrm{QCD}}}{m+\widetilde{m}}e^{\frac{m^2}{M^2}},
 \label{Eq:massResidue}
\end{eqnarray}
where $\mathcal{B}\widetilde{\Pi}
_{1(2)}^{\mathrm{QCD}}=-\frac{d}{d (1/M^2)}\mathcal{B}\Pi _{1(2)}^{\mathrm{QCD}}$.

The sum rules for  mass and residue of the radially excited state $\Xi'$
are obtained from Eq. (\ref{Eq:massResidue}) by replacements
$\widetilde{m}\rightarrow -m^{\prime}$ and 
$\widetilde{\lambda} \rightarrow \lambda^{\prime}$. Note that, there are other approaches to separate the contributions of the positive and
negative parity baryons (for instance see \cite{rev1,rev2,rev3,rev4}).

\begin{table}[tbp]
\begin{tabular}{|c|c|}
\hline\hline Parameters & Values \\ \hline\hline
$m_{s}$ & $96^{+8}_{-4}~\mathrm{GeV}$ \cite{PDG}\\
$m_{\Xi}$ & $(1314.86\pm0.20)~\mathrm{MeV}$ \cite{PDG}\\
$m_{\Sigma}$ & $(1189.37\pm0.07)~\mathrm{MeV}$ \cite{PDG}\\
$m_{\Lambda}$ & $(1115.683\pm0.006)~\mathrm{MeV}$ \cite{PDG}\\
$\lambda_{\Sigma}$ & $(0.014\pm0.03)~\mathrm{GeV^3}$ \cite{Aliev:2002ra}\\
$\lambda_{\Lambda}$ & $(0.013\pm0.02)~\mathrm{GeV^3}$ \cite{Aliev:2002ra}\\
$\langle \bar{u}u \rangle $ & $(-0.24\pm 0.01)^3$ $\mathrm{GeV}^3$  \\
$\langle \bar{s}s \rangle $ & $0.8\cdot(-0.24\pm 0.01)^3$ $\mathrm{GeV}^3$  \\
$\langle \overline{u}g_s\sigma Gu\rangle$ & $m_{0}^2\langle
\bar{u}u \rangle$
\\
$\langle \overline{s}g_s\sigma Gs\rangle$ & $m_{0}^2\langle
\bar{s}s \rangle$
\\
$m_{0}^2 $ & $(0.8\pm0.1)$ $\mathrm{GeV}^2$ \\
$\langle\frac{\alpha_sG^2}{\pi}\rangle $ & $(0.012\pm0.004)$ $~\mathrm{GeV}%
^4 $\\
\hline\hline
\end{tabular}%
\caption{Some input parameters used in the calculations.}
\label{tab:Param}
\end{table}

The sum rules for the mass and residue of the orbitally and
radially excited state of the $\Xi$ baryon as well as the residue
of the ground state contain many input parameters. Their values
are presented in Table~\ref{tab:Param}. For performing analysis of
widths of the $\Xi\rightarrow \Lambda K$ and $\Xi\rightarrow \Sigma K$
decays in next section, we also need the residues of the $\Lambda$ and $\Sigma$
baryons. We use the values of these residues
calculated via QCD sum rules 
\cite{Aliev:2002ra}. The mass of the ground state $\Xi$ is taken
as input parameter, as well. Besides these input parameters, QCD sum
rules contains three auxiliary parameters, namely the value of
continuum threshold $s_0$, Borel mass square $M^2$ and $\beta$
arbitrary parameter. Obviously any measurable physical quantity
must be independent of these parameters. Hence we need to find the
working regions of these parameters, where physical quantities
demonstrate good stability agains the variations of these parameters. The window for
$M^2$ is obtained by requiring that the series of operator product
expansion (OPE) in QCD side is convergent and the
contribution of higher states and continuum is sufficiently suppressed. Numerical analyses
lead to the conclusion that both conditions are satisfied in the
domain
\begin{eqnarray}
1.8\ \mathrm{GeV}^{2}\leq M^{2}& \leq &2.2\ \mathrm{GeV}^{2}.
\label{Eq:MsqMpsq}
\end{eqnarray}%
The considerations of the pole dominance and OPE convergence lead to the following
working window for the continuum threshold:
\begin{eqnarray}
1.9^2\,\,\mathrm{GeV}^{2}&\leq& s_{0}\leq 2.1^2\,\,\mathrm{GeV}^{2}.
\label{Eq:s0s0p}
\end{eqnarray}

In Figs.\ref{gr:mass1P}-\ref{gr:lam2S} we present, as examples, the dependence of
the  mass of the $\widetilde{\Xi}$ state and residues of the
$\Xi$, $\widetilde{\Xi}$ and $\Xi'$ baryons on $M^2$ and $ s_0 $ at
fixed value of $\cos\theta=0.7$, with $ \beta= \tan\theta$. From these figures we
observe that the results shows quite good stability with respect to the variations of
$M^2$ and $s_0$.

In order to find the working region of $\beta$, as an example in
Fig. \ref{gr:lamGorundCosTeta} we present the dependence of the
ground-state $\Xi$ baryon's residue on the $\cos\theta$. From this
figure we see that the residue is practically insensitive to the
variations of $\cos\theta$ in the domains
\begin{equation}
-1\leq \cos\theta \leq -0.3, \, \, \,0.3\leq \cos\theta \leq 1.
\end{equation}

\begin{widetext}

\begin{figure}[h!]
\begin{center}
\includegraphics[totalheight=6cm,width=8cm]{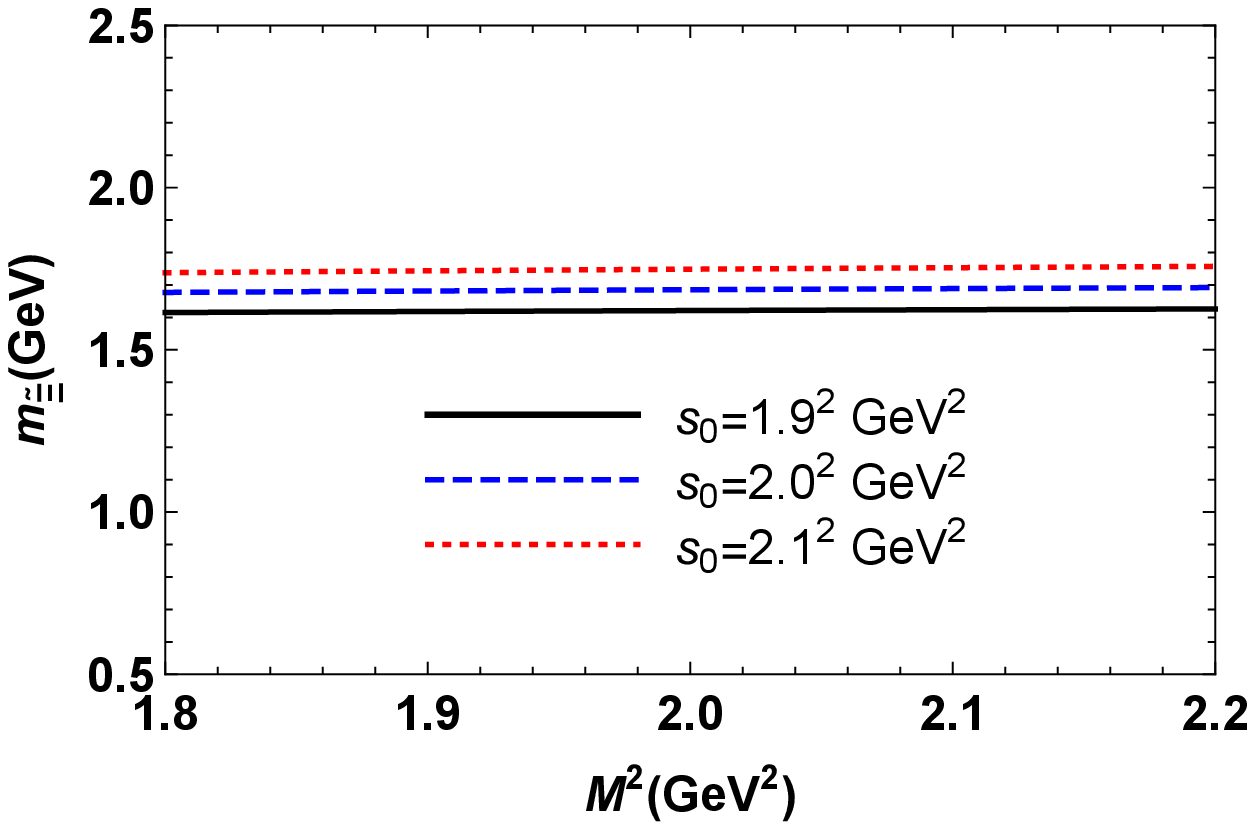}
\includegraphics[totalheight=6cm,width=8cm]{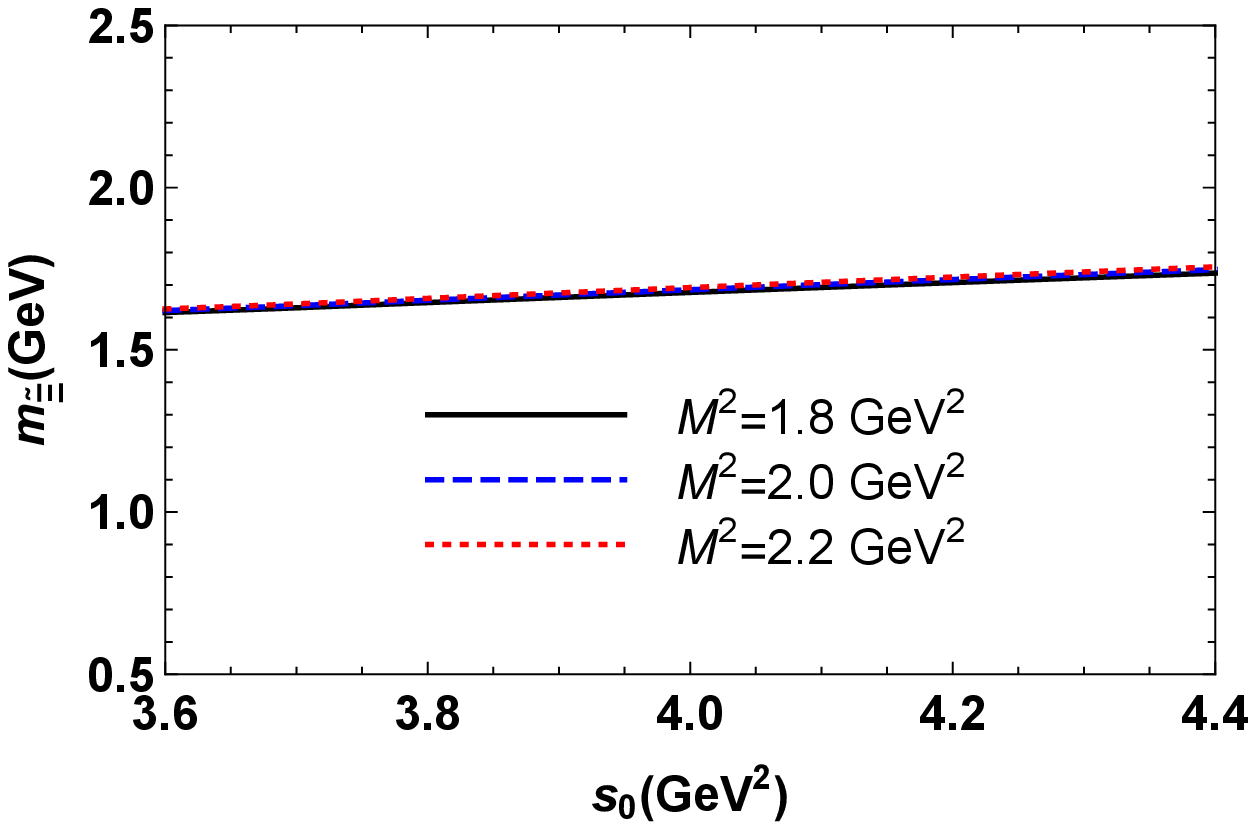}
\end{center}
\caption{ The mass of the $\widetilde{\Xi} $ baryon as a function
of the Borel parameter $M^2$ at chosen values of $s_0$ (left
panel), and as a function of the continuum threshold $s_0$ at
fixed $M^2$ (right panel) with $\beta=0.7$.} \label{gr:mass1P}
\end{figure}

\begin{figure}[h!]
\begin{center}
\includegraphics[totalheight=6cm,width=8cm]{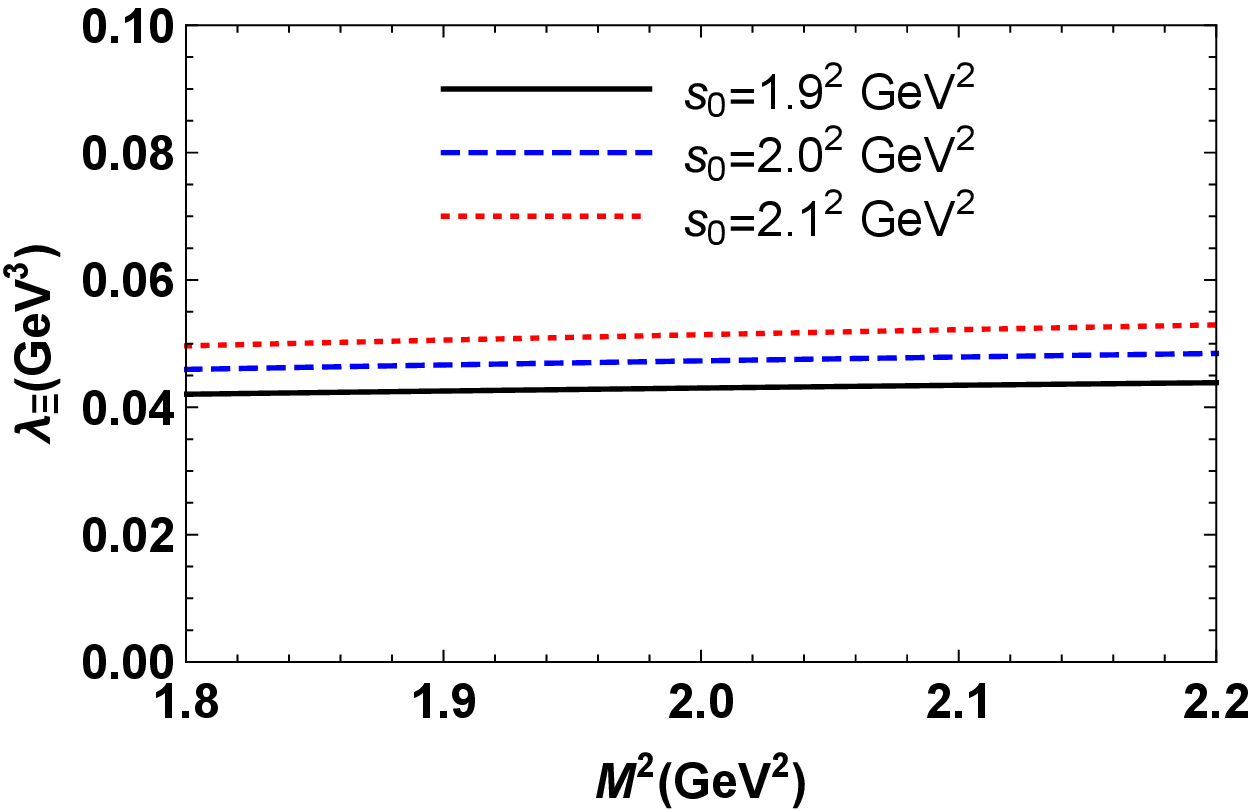}
\includegraphics[totalheight=6cm,width=8cm]{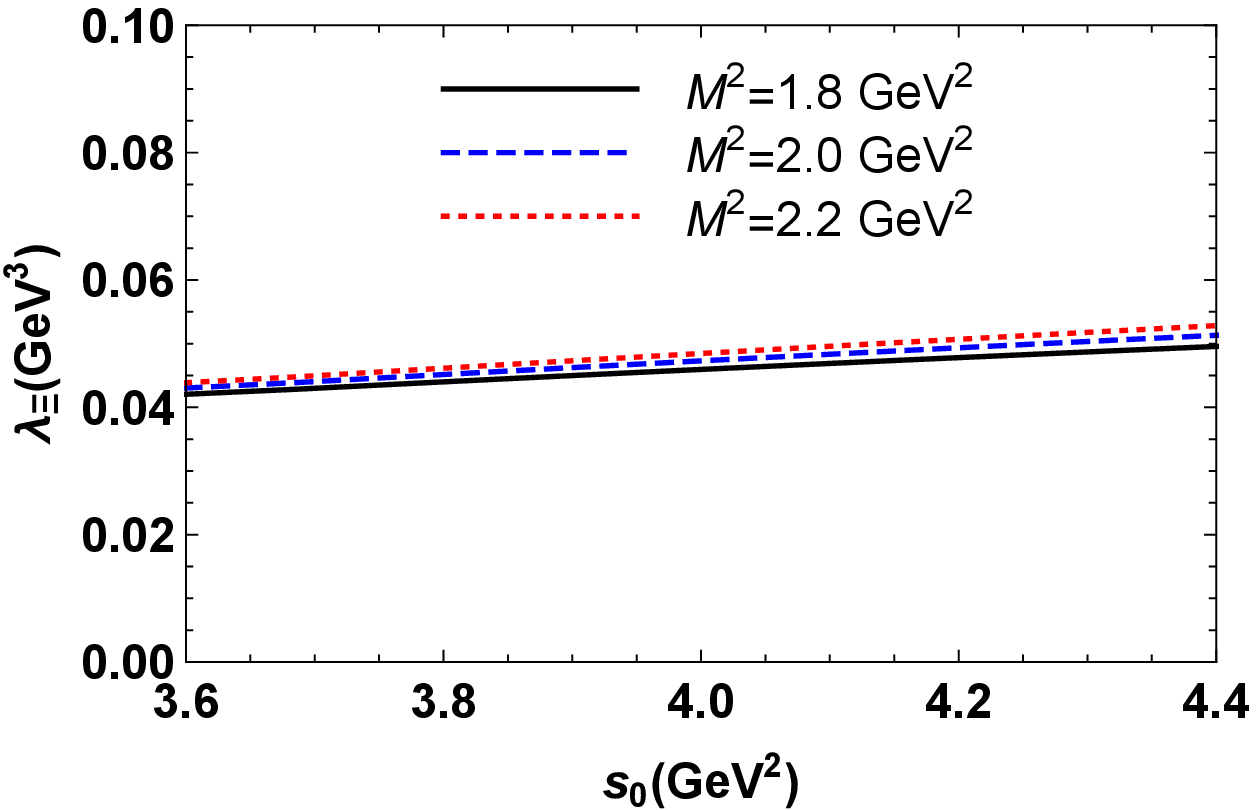}
\end{center}
\caption{ The rediue of the $\Xi$ baryon as a function of the
Borel parameter $M^2$ at chosen values of $s_0$ (left panel), and
as a function of the continuum threshold $s_0$ at fixed $M^2$
(right panel) with $\beta=0.7$.} \label{gr:GroundLam}
\end{figure}

\begin{figure}[h!]
\begin{center}
\includegraphics[totalheight=6cm,width=8cm]{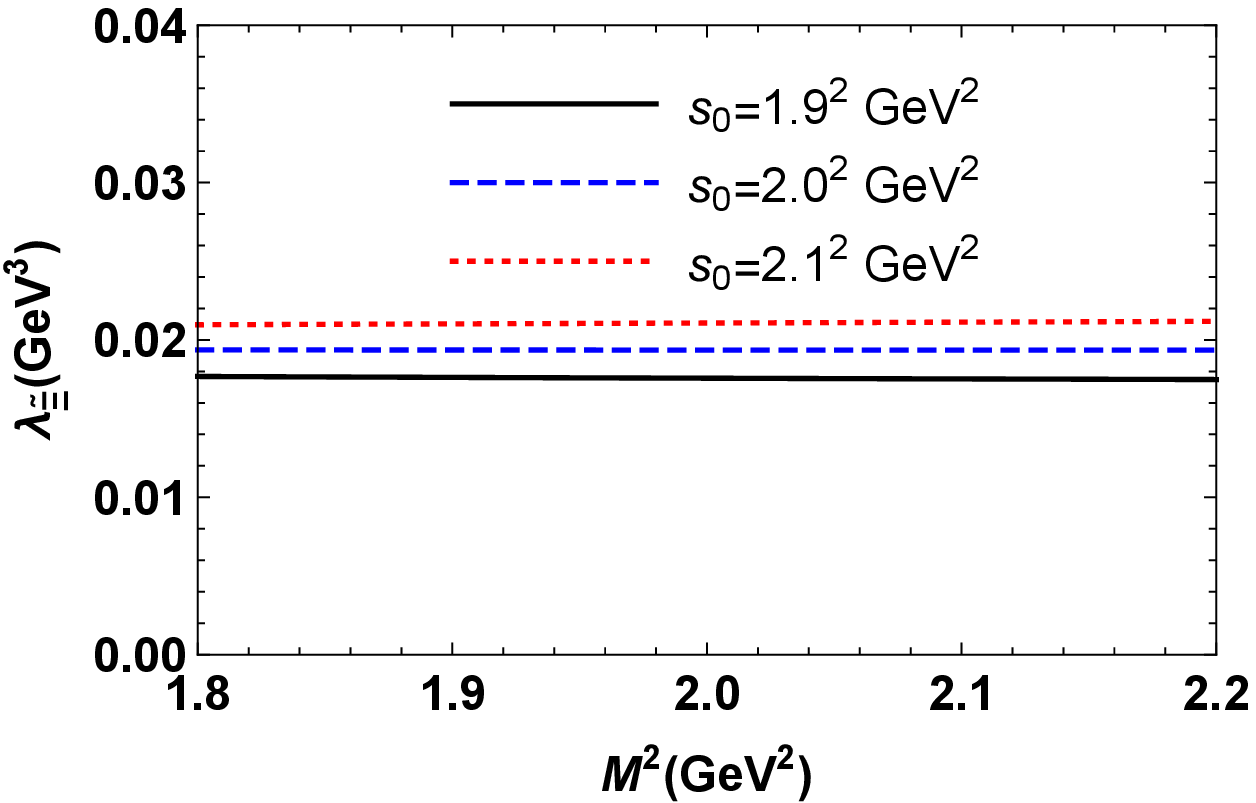}
\includegraphics[totalheight=6cm,width=8cm]{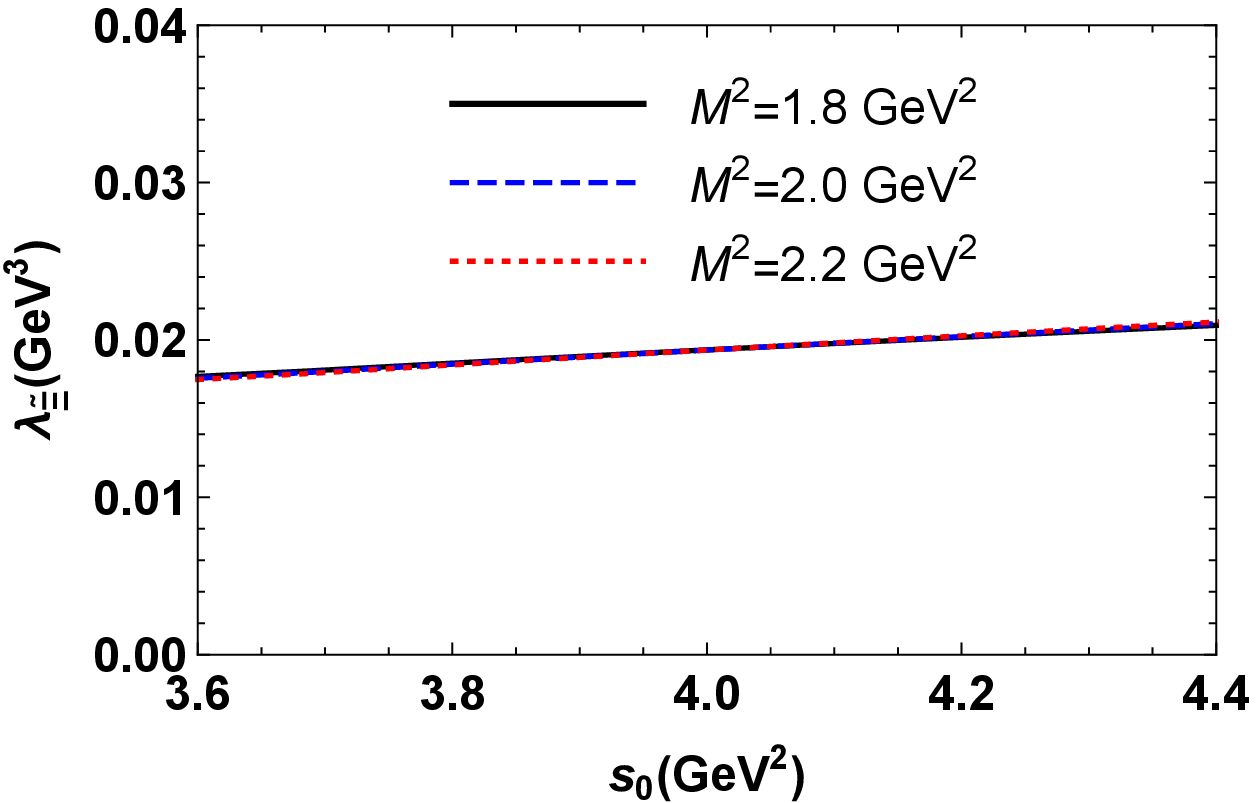}
\end{center}
\caption{The same as in Fig. \ref{gr:GroundLam}, but for the
orbitally excited $\widetilde{\Xi}$ baryon with $\beta=0.7$.}
\label{gr:lam1P}
\end{figure}

\begin{figure}[h!]
\begin{center}
\includegraphics[totalheight=6cm,width=8cm]{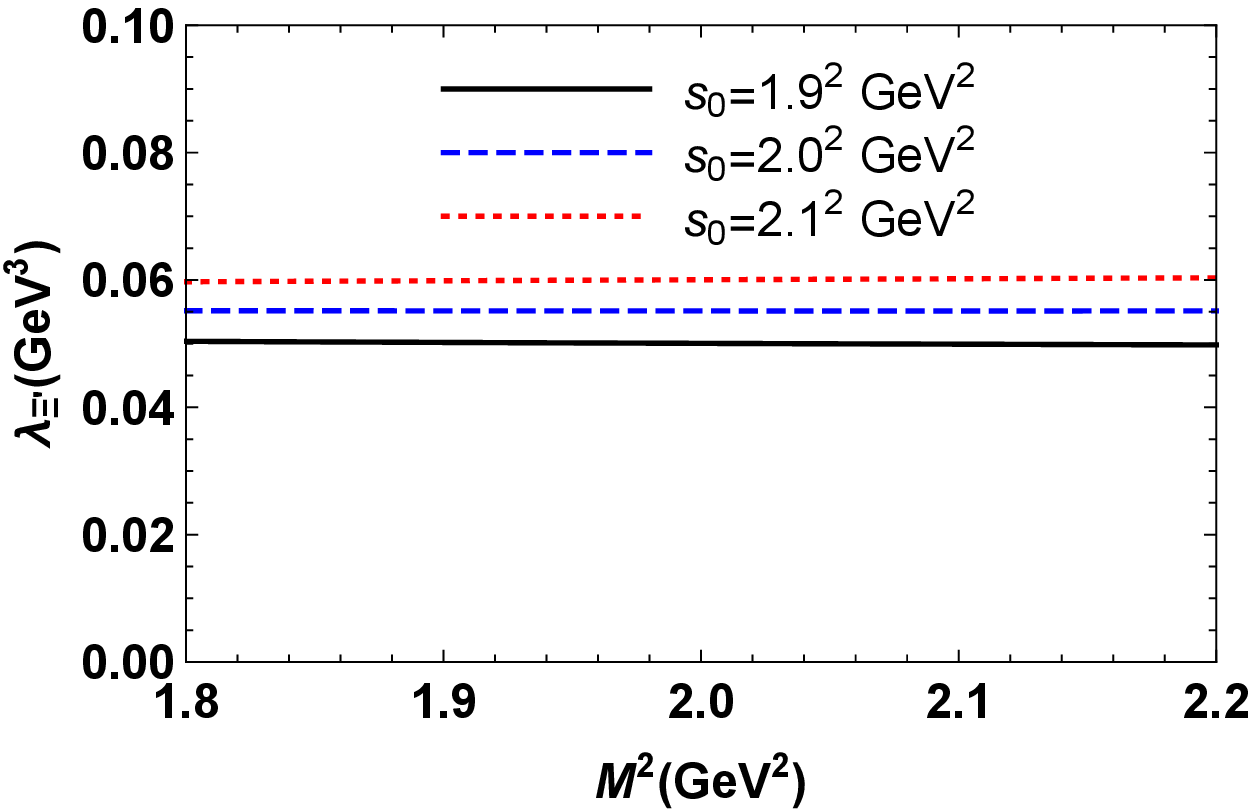}
\includegraphics[totalheight=6cm,width=8cm]{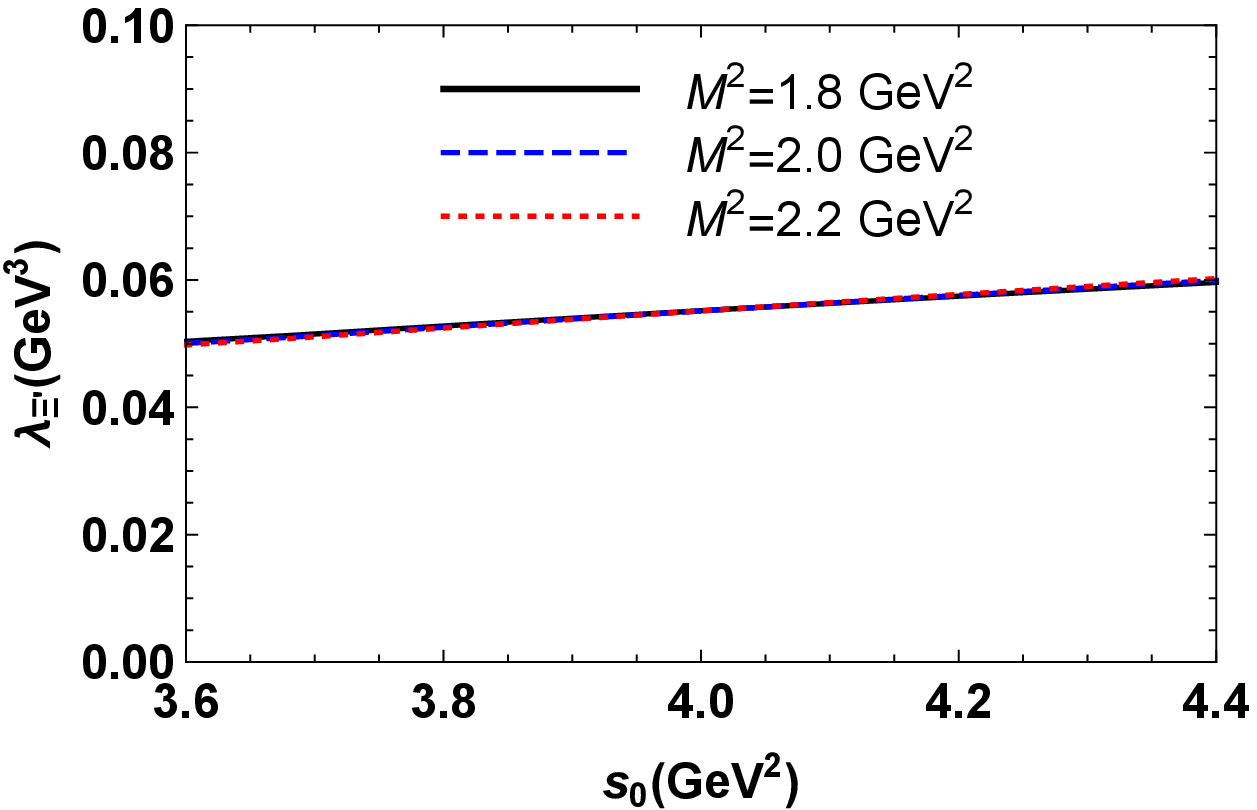}
\end{center}
\caption{The same as in Fig. \ref{gr:GroundLam}, but for the
radially excited $\Xi'$ baryon with $\beta=0.7$.} \label{gr:lam2S}
\end{figure}

\begin{figure}[h!]
\begin{center}
\includegraphics[totalheight=6cm,width=8cm]{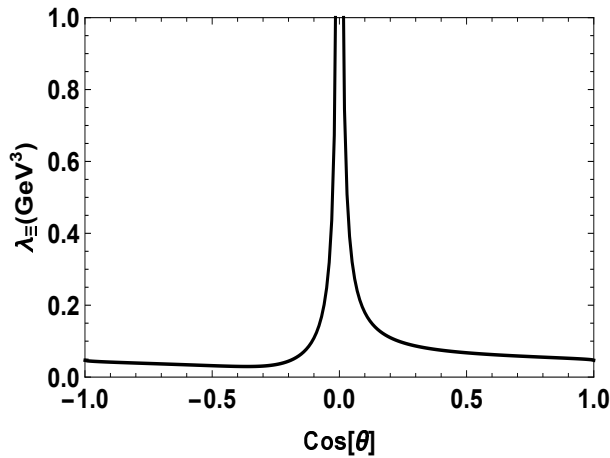}
\end{center}
\caption{The residue of the ground state $\Xi$ baryon as a
function of the $\cos\theta$ at central values of the $s_0$ and
$M^2$.} \label{gr:lamGorundCosTeta}
\end{figure}


\end{widetext}

We depict the numerical results of the masses and residues of the
first orbitally and radially excited $\Xi$ baryon as well as the ground state residue in table
\ref{tab:NumResults}. The errors in the presented results are due
to the uncertainties in determinations of the working regions of
the auxiliary parameters as well as the errors of other input
parameters. From this table we see that although consistent with the experimental data \cite{Aubert:2008ty,Abe:2001mb}, the radial and orbital excitation of $ \Xi $ receive  the same mass, which prevent us to assign any quantum numbers to  $ \Xi (1690)$ only via mass calculations. The residue of these two states are obtained to be differ from each other by a factor of roughly three.

\begin{table}[tbp]
\begin{tabular}{|c|c|c|c|}
\hline & $\Xi$ & $\widetilde{\Xi}$ & $\Xi'$\\ \hline \hline
$m~(\mathrm{MeV})$ &  & $1685\pm69$&$1685\pm69$ \\
\hline $\lambda  ~(\mathrm{GeV}^3)$ & $0.047\pm0.007$& $ 0.019\pm0.004$&$ 0.055
\pm0.010$ \\
\hline
\end{tabular}%
\caption{The sum rule results for the masses and residues of the
first orbitally and radially excited $\Xi$ baryon as well as
residue of the ground state.} \label{tab:NumResults}
\end{table}

\section{ $ \widetilde{\Xi}$ and $\Xi'$ Transitions To $\Lambda K$ and $\Sigma K$}

In present section we calculate the strong couplings
$g_{\widetilde{\Xi}\Sigma K}$, $g_{\widetilde{\Xi}\Lambda K}$,
$g_{\Xi'\Sigma K}$
 and $g_{\Xi'\Lambda K}$ defining the $\widetilde{\Xi}\rightarrow\Sigma
 K$,
 $\widetilde{\Xi}\rightarrow\Lambda K$, $\Xi' \rightarrow \Sigma
 K$ and $\Xi' \rightarrow \Lambda K$ transitions.

 For this aim we introduce the correlation
function
\begin{equation}
\Pi (p,q)=i\int d^{4}xe^{ipx}\langle K(q)|\mathcal{T}\{\eta
_{\Sigma(\Lambda) }(x) \bar{\eta }_{\Xi}(0)\}|0\rangle ,
\label{Eq:CorFunDecay}
\end{equation}%
where $\eta _{\Sigma}(x)$ and $\eta _{\Lambda}(x)$ are the
interpolating currents for the $\Sigma$ and $\Lambda$ baryons,
respectively. The general forms of these currents  are taken as \cite{Chung,Dosch}
\begin{eqnarray}\label{currents}
\eta_{\Sigma}(x)&=&-\frac{1}{\sqrt{2}}\epsilon_{abc}
\sum_{i=1}^{2}\Big[\Big(u^{T,a}(x)CA_{1}^{i}s^{b}(x)\Big)A_{2}^{i}d^{c}(x)
\nonumber \\
&-&\Big(d^{T,c}(x)CA_{1}^{i}s^{b}(x)\Big)A_{2}^{i}u^{a}(x)\Big],
\nonumber \\
\eta_{\Lambda}(x)&=&\frac{1}{\sqrt{6}}\epsilon_{abc}\sum_{i=1}^{2}
\Big[2\Big(u^{T,a}(x)CA_{1}^{i}
d^{b}(x)\Big)
A_{2}^{i}s^{c}(x)
\nonumber \\
&+&\Big(u^{T,a}(x)CA_{1}^{i}s^{b}(x)\Big)A_{2}^{i}d^{c}(x)\nonumber \\
&&+\Big(d^{T,c}(x)CA_{1}^{i}s^{b}(x)\Big)A_{2}^{i}u^{a}(x)\Big],
\end{eqnarray}
where $a, b, c$ are color indices, $C$ is the charge conjugation
operator and $A_{1}^{1}=I$, $A_{1}^{2}=A_{2}^{1}=\gamma_5$,
$A_{2}^{2}=\beta$. According to the method used, we again calculate the
aforesaid correlation function in two representations: hadronic and
QCD. Matching these two sides through a dispersion relation
leads to the sum rules for the coupling constants under consideration.

Firstly let us consider the $\widetilde{\Xi}\rightarrow \Sigma K$
transition. As we already noted,  the interpolating currents for
baryons can interact with both the positive and negative parity baryons. In what follows, we denote
the ground state positive (negative) parity baryons with 
$\Sigma(\widetilde{\Sigma})$ and $\Lambda(\widetilde{\Lambda})$.
Taking into account this fact,  inserting complete sets of
hadrons with the same quantum numbers as the interpolating
currents and isolating the ground states, we obtain

\begin{eqnarray}
\Pi ^{\mathrm{Phys}}(p,q)&=&\frac{\langle 0|\eta _{\Sigma}|\Sigma
(p,s)\rangle }{p^{2}-m_{\Sigma}^{2}}\langle K(q)\Sigma(p,s)|\Xi
(p^{\prime },s^{\prime })\rangle
\nonumber \\
&\times& \frac{\langle \Xi(p^{\prime },s^{\prime })|\bar{ \eta
}_{\Xi}|0\rangle }{p^{\prime 2}-m_{\Xi}^{2}}
\nonumber \\
&+& \frac{\langle 0|\eta_{\Sigma}|\widetilde{\Sigma} (p,s)\rangle
}{p^{2}-m_{\widetilde{\Sigma}}^{2}}\langle
K(q)\widetilde{\Sigma}(p,s)|\Xi
(p^{\prime },s^{\prime })\rangle  \nonumber \\
&\times& \frac{\langle \Xi(p^{\prime },s^{\prime })|\bar{\eta
}_{\Xi}|0\rangle }{p^{\prime 2}-m_{\Xi}^{2}}
\nonumber \\
&+&\frac{\langle 0|\eta _{\Sigma}|\Sigma (p,s)\rangle
}{p^{2}-m_{\Sigma}^{2}}\langle K(q)\Sigma(p,s)|\widetilde{\Xi}
(p^{\prime },s^{\prime })\rangle  \nonumber \\
&\times& \frac{\langle \widetilde{\Xi}(p^{\prime },s^{\prime
})|\bar{ \eta }_{\widetilde{\Xi}}|0\rangle }{p^{\prime
2}-m_{\widetilde{\Xi}}^{2}}
 \nonumber
\\&+&
\frac{\langle 0|\eta _{\Sigma}|\widetilde{\Sigma}(p,s)\rangle
}{p^{2}-m_{\widetilde{\Sigma}}^{2}}\langle K(q)\widetilde{\Sigma}
(p,s)|\widetilde{\Xi}
(p^{\prime },s^{\prime })\rangle  \nonumber \\
&\times& \frac{\langle \widetilde{\Xi} _{c}(p^{\prime },s^{\prime
})|\bar{\eta }_{\widetilde{\Xi}}|0\rangle }{p^{\prime
2}-m_{\widetilde{\Xi}}^{2}}
 +\ldots ,  \label{eq:SRDecay}
\end{eqnarray}
where $p^{\prime }=p+q,\ p$ and $q$ are the momenta of the $\Xi $,
$\Sigma$ baryons and $K$ meson, respectively.
 In this expression $
m_{\Sigma}$ is the mass of the $\Sigma$ baryon. The dots in Eq.\
(\ref{eq:SRDecay}) stand for contributions of the higher
resonances and continuum states.

The matrix elements in Eq. (\ref{eq:SRDecay}) are determined as
\begin{eqnarray}
\langle 0|\eta _{\Sigma }|\Sigma(p,s)\rangle & =&\lambda_{\Sigma}
u(p,s), \nonumber \\
\langle 0|\eta_{\widetilde{\Sigma}}|\widetilde{\Sigma}(p,s)\rangle
&=&\lambda_{\widetilde{\Sigma}}\gamma_5 u(p,s),
\nonumber \\
\langle K(q)\Sigma (p,s)|\Xi(p^{\prime },s^{\prime })\rangle &=&
g_{\Xi \Sigma K}\overline{u}(p,s)\gamma _{5}u(p^{\prime
},s^{\prime }),
\nonumber \\
\langle K(q)\Sigma (p,s)|\widetilde{\Xi} (p^{\prime },s^{\prime
})\rangle &=& g_{\widetilde{\Xi} \Sigma
K}\overline{u}(p,s)u(p^{\prime },s^{\prime }),
\nonumber \\
\langle K(q)\widetilde{\Sigma} (p,s)|\Xi(p^{\prime },s^{\prime
})\rangle &=& g_{\Xi\widetilde{\Sigma} K}\overline{u}(p,s)
u(p^{\prime },s^{\prime }),
\nonumber \\
\langle K(q)\widetilde{\Sigma} (p,s)|\widetilde{\Xi}(p^{\prime
},s^{\prime })\rangle &=&g_{\widetilde{\Xi}\widetilde{\Sigma}
K}\overline{u}(p,s)\gamma_5 u(p^{\prime },s^{\prime }). \nonumber
\\  \label{eq14}
\end{eqnarray}
where $g_i$ are the strong coupling constants for the corresponding
transitions.

Using the matrix elements given in Eq.(\ref{eq14}) and performing
summation over spins of $\Sigma$ and $\Xi$ baryons and applying
the double Borel transformations with respect $p^2$ and $p^{\prime
2}$ for physical side of the correlation function we get
\begin{eqnarray}
&&\mathcal{B}\Pi ^{\mathrm{Phys}}(p,q)=g_{\Xi \Sigma K}\lambda
_{\Xi}\lambda_{\Sigma}e^{-m_{\Xi}^{2}/M_{1}^{2}}e^{-m_{\Sigma}^{2}/M_{2}^{2}}  \notag \\
&&\times \left( \slashed p+m_{\Sigma}\right)\gamma_5\left(
\slashed p+\slashed q+m_{\Xi}\right)   \notag \\
&&-g_{\widetilde{\Xi} \widetilde{\Sigma} K}\lambda
_{\widetilde{\Xi}
}\lambda_{\widetilde{\Sigma}}e^{-m_{\widetilde{\Xi}}^{2}/M_{1}^{2}}
e^{-m_{\widetilde{\Sigma}}^{2}/M_{2}^{2}}  \notag \\
&&\times \gamma_5\left( \slashed
p+m_{\widetilde{\Sigma}}\right)\gamma_5\left( \slashed p+\slashed
q+m_{\widetilde{\Xi}}\right)\gamma_5
\notag \\
&&-g_{\widetilde{\Xi} \Sigma K}\lambda
_{\widetilde{\Xi}}\lambda_{\Sigma}e^{-m_{\widetilde{\Xi}}^{2}/M_{1}^{2}}
e^{-m_{\Sigma}^{2}/M_{2}^{2}}  \notag \\
&&\times \left( \slashed p+m_{\Sigma}\right)\left( \slashed
p+\slashed q+m_{\widetilde{\Xi}}\right)\gamma_5
\notag \\
&&+ g_{\Xi \widetilde{\Sigma} K}\lambda
_{\Xi}\lambda_{\widetilde{\Sigma}}e^{-m_{\Xi}^{2}/M_{1}^{2}}e^{-m_{
\widetilde{\Sigma}}^{2}/M_{2}^{2}}  \notag \\
&&\times \gamma_5 \left( \slashed
p+m_{\widetilde{\Sigma}}\right)\left( \slashed p+\slashed
q+m_{\Xi}\right) , \label{eq:CFunc1/2}
\end{eqnarray}%
where  $M_{1}^{2}$ and $
M_{2}^{2}$ are the Borel parameters.

From Eq. (\ref{eq:CFunc1/2}) it follows that we have different
structures  which can be used to obtain sum rules for the
strong coupling constant of $\widetilde{\Xi}\rightarrow \Sigma K$
channel. We have four couplings (see Eq.\ref{eq:CFunc1/2}), and in
order to determine the coupling $g_{\widetilde{\Xi}\Sigma K}$ we
need four equations. Therefore we select the structures $\slashed
q\slashed p \gamma_5$, $\slashed p \gamma_5$, $\slashed q
\gamma_5$ and $\gamma_5$. Solving four algebraic equations for
$g_{\widetilde{\Xi}\Sigma K}$, finally we get
\begin{eqnarray}
g_{\widetilde{\Xi} \Sigma
K}&=&\frac{e^{\frac{m_{\widetilde{\Xi}}^2}{M_1^2}}e^{\frac{m_{\Sigma
}^2}{M_2^2}}}
{\lambda_{\widetilde{\Xi}}\lambda_{\Sigma}(m_{\Sigma}
+m_{\widetilde{\Sigma}})
(m_{\Xi}+m_{\widetilde{\Xi}})} \nonumber \\
&\times& \left[\Pi
_{1}^{\mathrm{^{\prime}OPE}}\left(m_K^2+m_{\widetilde{\Sigma}}m_{\Xi}-m_{\Xi}
m_{\widetilde{\Xi}}\right)\right.
\nonumber \\
&+&\left. \Pi
_{2}^{\mathrm{^{\prime}OPE}}\left(m_{\widetilde{\Xi}}-m_{\widetilde{\Sigma}}
-m_{\Xi}\right) \right.
\nonumber \\
&+&\left. \Pi
_{3}^{\mathrm{^{\prime}OPE}}\left(m_{\widetilde{\Sigma}}-m_{\widetilde{\Xi}}
\right)-\Pi _{4}^{\mathrm{^{\prime}OPE}}\right],\label{eq16}
\end{eqnarray}
where $\Pi _{1}^{\mathrm{^{\prime}OPE}}$, $\Pi
_{2}^{\mathrm{^{\prime}OPE}}$, $\Pi _{3}^{\mathrm{^{\prime}OPE}}$
and $\Pi _{4}^{\mathrm{^{\prime}OPE}}$ are the invariant
amplitudes corresponding to the structures $\slashed q\slashed
p\gamma _{5}$, $\slashed p\gamma _{5}$, $\slashed q\gamma _{5}$
and $\gamma _{5}$ for $\widetilde{\Xi}\rightarrow \Sigma K$ decay,
respectively.

If we carry out the same procedures for $\widetilde{\Xi}\rightarrow
\Lambda K$ decay, for the coupling constant $g_{\widetilde{\Xi}\Lambda
K}$ we obtain:

\begin{eqnarray}
g_{\widetilde{\Xi} \Lambda
K}&=&\frac{e^{\frac{m_{\widetilde{\Xi}}^2}{M_1^2}}e^{\frac{m_{\Lambda
}^2}{M_2^2}}}
{\lambda_{\widetilde{\Xi}}\lambda_{\Lambda}(m_{\Lambda}
+m_{\widetilde{\Lambda}})
(m_{\Xi}+m_{\widetilde{\Xi}})} \nonumber \\
&\times& \left[\widetilde{\Pi}
_{1}^{\mathrm{^{\prime}OPE}}\left(m_K^2+m_{\widetilde{\Lambda}}m_{\Xi}-m_{\Xi}
m_{\widetilde{\Xi}}\right)\right.
\nonumber \\
&+&\left. \widetilde{\Pi}
_{2}^{\mathrm{^{\prime}OPE}}\left(m_{\widetilde{\Xi}}-m_{\widetilde{\Lambda}}
-m_{\Xi}\right) \right.
\nonumber \\
&+&\left. \widetilde{\Pi}
_{3}^{\mathrm{^{\prime}OPE}}\left(m_{\widetilde{\Lambda}}-m_{\widetilde{\Xi}}
\right)-\widetilde{\Pi}
_{4}^{\mathrm{^{\prime}OPE}}\right],\label{eq16a}
\end{eqnarray}
where $\widetilde{\Pi}_{1}^{\mathrm{^{\prime}OPE}}$,
$\widetilde{\Pi}_{2}^{\mathrm{^{\prime}OPE}}$, $\widetilde{\Pi}
_{3}^{\mathrm{^{\prime}OPE}}$ and $\widetilde{\Pi}
_{4}^{\mathrm{^{\prime}OPE}}$ are the invariant amplitudes
corresponding to the structures $\slashed q\slashed p\gamma _{5}$,
$\slashed p\gamma _{5}$, $\slashed q\gamma _{5}$ and $\gamma _{5}$
for $\widetilde{\Xi}\rightarrow \Lambda K$ decay, respectively.

The general expressions obtained above contain two Borel
parameters $M_1^{2}$ and $M_1^{2}$. In our analysis we choose
\begin{equation}
M_{1}^{2}=M_{2}^{2}= 2M^{2},\,\,
M^2=\frac{M_1^{2}M_2^{2}}{M_1^{2}+M_2^{2}},
\end{equation}
since the masses of the involved  $\Xi$ and $\Sigma (\Lambda)$ are
close to each other.

The sum rules for the coupling constants for $\Xi' \rightarrow \Sigma
K$ and $\Xi' \rightarrow \Lambda K$ transitions can be easily
obtained from Eqs. (\ref{eq16}) and (\ref{eq16a}), by replacing
$m_{\widetilde{\Xi}}\rightarrow - m_{\Xi^{\prime}}$ and
$\lambda_{\widetilde{\Xi}}\rightarrow \lambda_{\Xi^{\prime}}$.

The OPE side of the correlation function  $\Pi
^{\mathrm{OPE}}(p,q)$ can be obtained by inserting the corresponding interpolating currents to the correlation function, using Wick's theorem to contract the quark fields, and inserting into the obtained expression the relevant
quark propagators. The nonperturbative contributions in
light cone QCD sum rules, which are described in terms of the $K$-meson
distribution amplitudes (DAs), can be obtained by using Fierz rearrangement
formula
\begin{equation*}
\overline{s}_{\alpha }^{a}u_{\beta }^{b}=\frac{1}{4}\Gamma _{\beta
\alpha }^{i}(\overline{s}^{a}\Gamma ^{i}u^{b}),
\end{equation*}%
where $\ \Gamma ^{i}=1,\ \gamma _{5},\ \gamma _{\mu },\ i\gamma
_{5}\gamma _{\mu },\ \sigma _{\mu \nu }/\sqrt{2}$ is the full set
of Dirac matrices. The matrix elements of these terms between the
$K$-meson and vacuum states, as well as ones generated by insertion
of the gluon field strength tensor $G_{\lambda \rho }(uv)$ \ from
quark propagators, are determined in terms of the $K$-meson
DAs  with definite twists.
The DAs are main nonperturbative inputs of light cone QCD sum
rules. The $K$-meson distribution amplitudes are derived in
\cite{Ball:2006wn,Belyaev:1994zk,Ball:2004ye} which will be used in
our numerical analysis.  All of these steps summarized above result
in lengthy expression for the OPE side of correlation function. In
order not to overwhelm the study with overlong mathematical
expressions we prefer not to present them here. Apart from
parameters in the distribution amplitudes, the sum rules for the
couplings depend also on numerical values of the $\Sigma$ and
$\Lambda$ baryon's mass and pole residue, which are given in Table I.
Note that the working region of the
Borel mass $M^2$, threshold $s_0$ and $\beta$ parameters for
calculations of the relevant couplings are chosen the same as in the residue
and mass computations.

Performing numerical analysis for the relevant coupling constants we
get values presented in Table \ref{tab:NumResults1}.
Using the couplings $g_{\widetilde{\Xi} \Sigma K}$, $g_{\Xi'
\Sigma K}$ $g_{\widetilde{\Xi} \Lambda K}$ and $g_{\Xi' \Lambda
K}$ we can easily calculate the width of $\widetilde{\Xi}
\rightarrow \Sigma K$, $\Xi' \rightarrow \Sigma K$,
$\widetilde{\Xi} \rightarrow \Lambda K$ and $\Xi' \rightarrow
\Lambda K$
decays. After some computations we obtain:%
\begin{eqnarray}
\Gamma \left( \widetilde{\Xi} \rightarrow \Sigma K\right)
&=&\frac{ g_{\widetilde{\Xi} \Sigma K}^{2}}{16\pi
m_{\widetilde{\Xi}}^{3}}\left[ (m_{\widetilde{\Xi}}
+m_{\Sigma})^{2}-m_{K}^{2}\right]  \notag \\
&&\times \lambda^{1/2}(m_{\widetilde{\Xi}}^2,m_{\Sigma
}^2,m_{K}^2), \label{DW1}
\end{eqnarray}
and

\begin{eqnarray}
\Gamma \left( \Xi^{\prime }\rightarrow \Sigma K\right) &=&%
\frac{g_{\Xi ^{\prime }\Sigma K}^{2}}{16\pi m_{\Xi'}^{
3}}\left[ (m_{\Xi'}-m_{\Sigma})^{2}-m_{K}^{2}\right]  \notag \\
&&\times \lambda^{1/2}(m_{\Xi'}^2,m_{\Sigma}^2,m_{K}^2).
\label{DW2}
\end{eqnarray}
In expressions above the function $\lambda(x^2,y^2,z^2)$ is given
as:
\begin{equation*}
\lambda(x^2,y^2,z^2)=
x^{4}+y^{4}+z^{4}-2x^{2}y^{2}-2x^{2}z^{2}-2y^{2}z^{2}.
\end{equation*}

The expressions for the  widths of the  $\widetilde{\Xi} \rightarrow \Lambda K$
and $\Xi^{\prime } \rightarrow \Lambda K$  can be easily obtained
from Eqs. (\ref{DW1}) and(\ref{DW2}), by the replacement
$m_{\Sigma}\rightarrow m_{\Lambda}$.

Using the values of coupling constants and formulas for the decay widths  we obtain the values of the partial width at  different decay channels presented in Table
\ref{tab:NumResults1}.
\begin{table}[tbp]
\begin{tabular}{|c|c|c|}
\hline &$ g$& $\Gamma (\mathrm{MeV})$\\
\hline \hline
$\widetilde{\Xi} \rightarrow \Sigma K$ &$1.35\pm0.37$  & $32.73\pm9.16$ \\
\hline $\Xi^{\prime }\rightarrow \Sigma K$ & $69.57\pm19.48$& $
3.08\pm0.86$
 \\
 \hline
$\widetilde{\Xi} \rightarrow \Lambda K$ &$1.65\pm0.48$  & $65.13\pm18.89$ \\
\hline $\Xi^{\prime }\rightarrow \Lambda K$ & $8.41\pm2.44$& $
18.23\pm5.29$
 \\
\hline
\end{tabular}%
\caption{The sum rule results for the strong coupling constants
and decay widths of the first orbitally and radially excited $\Xi$
baryon.} \label{tab:NumResults1}
\end{table}
Using the values of the partial decay widths from Table
\ref{tab:NumResults1}, we finally obtain the ratio
of the branching fractions in $\widetilde{\Xi}$ channel  as
\begin{equation}
\frac{Br\Big(\widetilde{\Xi} \rightarrow \Sigma K^-
\Big)}{Br\Big(\widetilde{\Xi} \rightarrow \Lambda \overline{K}^0
\Big)}=0.50\pm0.14, \label{RatioBRNum}
\end{equation}
and for $\Xi^{\prime}$ channel we get
\begin{equation}
\frac{Br\Big(\Xi^{\prime} \rightarrow \Sigma K^-
\Big)}{Br\Big(\Xi^{\prime} \rightarrow \Lambda \overline{K}^0
\Big)}=0.17\pm0.05. \label{RatioBRNum}
\end{equation}

As is seen, the obtained value for the ratio of the branching fractions in
$\widetilde{\Xi}$ channel is in nice consistency with the
experimental data of Belle Collaboration \cite{Abe:2001mb}:
\begin{equation}
\frac{Br\Big(\Xi(1690) \rightarrow \Sigma^+ K^-
\Big)}{Br\Big(\Xi(1690) \rightarrow \Lambda^0 \overline{K}^0
\Big)}=0.50\pm0.26, \label{RatioBRExp}
\end{equation}
Note that in \cite{Khemchandani:2017wfb}, within the coupled channel
approach, a very similar results has been found. The authors have concluded that the $\Xi(1690)$ has spin-$1/2$, but
its parity has not been established.
Our prediction for the corresponding ratio in $\Xi^{\prime }$ channel is considerably
small compared to the experimental  data. From these
results and those for the values of the corresponding masses we conclude that the $\Xi(1690)$ state, most probably, has
quantum numbers $\frac{1}{2}^{-}$, i.e. it represents  a negative
parity spin-1/2 baryon.

\section{ Acknowledgments}

K. A. thanks Dogus University for the partial financial support through the grant BAP 2015-16-D1-B04.

\appendix*


\section{ The QCD side of the correlation function in mass sum rules}

\label{sec:App}
\renewcommand{\theequation}{\Alph{section}.\arabic{equation}}

In present Appendix we present explicit forms of the functions in QCD side of the
two point correlation function used in mass sum rules:

\begin{widetext}

\begin{eqnarray}
\mathcal{B}\Pi^{\mathrm{QCD}}_1(q)&=&\int_{0}^{s_0}ds~
e^{-\frac{s}{M^2}}
\frac{1}{2^7\pi^2}\Big\{\frac{s^2(5\beta^2+2\beta+5)}{2^4\pi^2}+m_s
\Big[\langle \bar{s}s\rangle(5\beta^2+2\beta+5)+6\Big(\langle
\bar{u}u\rangle+\langle \bar{d}d\rangle\Big)(\beta^2-1)\Big]
\nonumber \\
&+&  \frac{\langle g_s^2 GG\rangle
(5\beta^2+2\beta+5)}{2^4\pi^2}-\frac{3m_0^2\Big(\langle
\bar{u}u\rangle+\langle
\bar{d}d\rangle\Big)m_s(\beta^2-1)}{M^2}\log\left[\frac{s}{\Lambda^2}
\right]+\frac{\langle g_s^2 GG\rangle\Big(\langle
\bar{u}u\rangle+\langle \bar{d}d\rangle\Big)}{3M^4}
\nonumber \\
&\times& m_s(\beta^2-1)\log\left[\frac{s}{\Lambda^2} \right]
\Big\}-\frac{m_s}{3\cdot2^8\pi^2}\Big[3m_0^2\Big(\langle
\bar{u}u\rangle+\langle
\bar{d}d\rangle\Big)(\beta^2-1)(6\gamma_E-13)-8m_0^2\langle
\bar{s}s\rangle(\beta^2+\beta+1)\Big]
\nonumber \\
&+&\frac{\beta-1}{24}\Big[3\Big(\langle \bar{s}s\rangle\langle
\bar{d}d\rangle+\langle \bar{s}s\rangle\langle
\bar{u}u\rangle\Big)(\beta+1)+\langle \bar{u}u\rangle\langle
\bar{d}d\rangle(\beta-1)\Big]+\frac{m_s(\beta+1)}{3\cdot2^9\pi^2M^2}\langle
g_s^2 GG\rangle\Big[4\Big(\langle \bar{u}u\rangle+\langle
\bar{d}d\rangle\Big)
\nonumber \\
&\times& (\beta-1)-\langle
\bar{s}s\rangle(\beta+1)\Big]+e^{-\frac{s_0}{M^2}}\frac{\langle
g_s^2 GG\rangle\Big(\langle \bar{u}u\rangle+\langle
\bar{d}d\rangle\Big)}{3\cdot2^7\pi^2s_0M^2}m_s(\beta^2-1)\Big(M^2+s_0\log
\Big[\frac{s_0}{\Lambda^2}\Big]\Big)
\nonumber \\
&+&\frac{(\beta-1)}{3\cdot2^5M^2}m_0^2\Big[6\Big(\langle
\bar{s}s\rangle\langle \bar{d}d\rangle+\langle
\bar{s}s\rangle\langle \bar{u}u\rangle\Big)+(\beta-1)\langle
\bar{u}u\rangle\langle
\bar{d}d\rangle\Big]+\frac{m_s(\beta^2-1)}{3\cdot2^(11)\pi^2M^4}\langle
g_s^2 GG\rangle m_0^2\Big(\langle \bar{u}u\rangle+\langle
\bar{d}d\rangle\Big),
\nonumber \\
\end{eqnarray}
and
\begin{eqnarray}
\mathcal{B}\Pi^{\mathrm{QCD}}_2(q)&=&\int_{0}^{s_0}ds~
e^{-\frac{s}{M^2}} \frac{1}{2^6\pi^2}\Big\{\frac{m_s
s^2(\beta-1)^2}{2^3\pi^2}+s(\beta-1)\Big[3\Big(\langle
\bar{u}u\rangle+\langle \bar{d}d\rangle\Big)(\beta+1)+\langle
\bar{s}s\rangle(\beta-1)\Big]
\nonumber \\
&-& \frac{\langle g_s^2
GG\rangle}{3\cdot2^4\pi^2}m_s(\beta-1)^2\Big(8+3\gamma_E-3\log\Big[
\frac{s}{\Lambda^2}\Big]\Big)+\frac{3m_0^2\Big(\langle
\bar{u}u\rangle+\langle
\bar{d}d\rangle\Big)}{2}(\beta^2-1)\Big\}+\frac{\langle g_s^2
GG\rangle}{2^{10}\pi^4}
\nonumber \\
&\times&\gamma_Em_sM^2(\beta-1)^2\Big(1-e^{-\frac{s_0}{M^2}}
\Big)-\frac{m_s}{3\cdot2^4}\Big[3\Big(\langle
\bar{s}s\rangle\langle \bar{d}d\rangle+\langle
\bar{s}s\rangle\langle \bar{u}u\rangle\Big)(\beta^2-1)+2\langle
\bar{u}u\rangle\langle \bar{d}d\rangle
\nonumber \\
&\times&
(5\beta^2+2\beta+5)\Big]+\frac{\beta-1}{3\cdot2^8\pi^2}\langle
g_s^2 GG\rangle\Big[3\Big(\langle \bar{u}u\rangle+\langle
\bar{d}d\rangle\Big)(\beta+1)-\langle
\bar{s}s\rangle(\beta-1)\Big]+\frac{\langle g_s^2 GG\rangle\langle
\bar{u}u\rangle\langle \bar{d}d\rangle }{3^2\cdot2^6M^4}
\nonumber \\
&\times&m_s(5\beta^2+2\beta+5)+\frac{m_0^2\langle
\bar{u}u\rangle\langle
\bar{d}d\rangle}{3\cdot2^4M^2}m_s(3\beta^2+2\beta+3)+\frac{\langle
g_s^2 GG\rangle m_0^2\langle \bar{u}u\rangle\langle
\bar{d}d\rangle}{3^2\cdot2^6M^6}m_s(5\beta^2+2\beta+5),
\nonumber \\
\end{eqnarray}
where, to shorten the expressions,   the terms proportional to $m_u$ and $m_d$ are not
presented, although their contributions are taken into account in
performing numerical analysis.

\end{widetext}
%



\begin{thebibliography}{999}



\bibitem{Isgur:1978xj}
  N.~Isgur and G.~Karl,
  Phys.\ Rev.\ D {\bf 18}, 4187 (1978).

\bibitem{Oh:2007cr}
Y.~Oh,
Phys.\ Rev.\ D {\bf 75}, 074002 (2007)  [hep-ph/0702126 [hep-ph]].

\bibitem{Lee:2002jb}
F.~X.~Lee and X.~Y.~Liu,
Phys.\ Rev.\ D {\bf 66}, 014014 (2002)  [nucl-th/0203051].

\bibitem{Pervin:2007wa}
  M.~Pervin and W.~Roberts,
Phys.\ Rev.\ C {\bf 77}, 025202 (2008)  [arXiv:0709.4000
[nucl-th]].



\bibitem{Melde:2008yr}
  T.~Melde, W.~Plessas and B.~Sengl,
Phys.\ Rev.\ D {\bf 77}, 114002 (2008)  [arXiv:0806.1454
[hep-ph]].

\bibitem{Schat:2001xr}
  C.~L.~Schat, J.~L.~Goity and N.~N.~Scoccola,
Phys.\ Rev.\ Lett.\  {\bf 88}, 102002 (2002)  [hep-ph/0111082].

\bibitem{Xiao:2013xi}
  L.~Y.~Xiao and X.~H.~Zhong,
  Phys.\ Rev.\ D {\bf 87}, no. 9, 094002 (2013)
  [arXiv:1302.0079 [hep-ph]].



\bibitem{Chao:1980em}
  K.~-T.~Chao, N.~Isgur and G.~Karl,
Phys.\ Rev.\ D {\bf 23}, 155 (1981).



\bibitem{Capstick:1986bm}
  S.~Capstick and N.~Isgur,
Phys.\ Rev.\ D {\bf 34}, 2809 (1986).



\bibitem{Aubert:2008ty}
  B.~Aubert {\it et al.} [BaBar Collaboration],
  Phys.\ Rev.\ D {\bf 78}, 034008 (2008)
  [arXiv:0803.1863 [hep-ex]]; B. Aubert et al. [BaBar Collaboration], hep-ex/0607043; V. Ziegler, SLAC-R-868..

\bibitem{Abe:2001mb}
  K.~Abe {\it et al.} [Belle Collaboration],
  Phys.\ Lett.\ B {\bf 524}, 33 (2002)
  [hep-ex/0111032].







\bibitem{Chung} V. Chung, H. G. Dosch, M. Kremer, D. Scholl, Nucl. Phys. B {\bf 197}, 55 (1982).

\bibitem{Dosch} H. G. Dosch, M. Jamin and S. Narison, Phys. Lett. B {\bf 220}, 251 (1989).

\bibitem{rev1} E. Bagan, M. Chabab, H. G. Dosch and S. Narison, Phys. Lett. B {\bf 301}, 243 (1993).

\bibitem{rev2} D. Jido, N. Kodama and M. Oka, Phys. Rev. D {\bf 54}, 4532 (1996).

\bibitem{rev3} Z. G. Wang, Phys. Lett. B {\bf 685}, 59 (2010).

\bibitem{rev4} Z. G. Wang, Eur. Phys. J. A {\bf 45}, 267 (2010).

\bibitem{PDG}
C. Patrignani et al. (Particle Data Group), Chin. Phys. C, {\bf
40}, 100001 (2016) and 2017 update.

\bibitem{Aliev:2002ra}
  T.~M.~Aliev, A.~Ozpineci and M.~Savci,
  Phys.\ Rev.\ D {\bf 66}, 016002 (2002)
  Erratum: [Phys.\ Rev.\ D {\bf 67}, 039901 (2003)]
  [hep-ph/0204035].


\bibitem{Ball:2006wn}
  P.~Ball, V.~M.~Braun and A.~Lenz,
  JHEP {\bf 0605}, 004 (2006)

\bibitem{Belyaev:1994zk}
  V.~M.~Belyaev, V.~M.~Braun, A.~Khodjamirian and R.~Ruckl,
  Phys.\ Rev.\ D {\bf 51}, 6177 (1995)



\bibitem{Ball:2004ye}
  P.~Ball and R.~Zwicky,
  Phys.\ Rev.\ D {\bf 71}, 014015 (2005)



\bibitem{Khemchandani:2017wfb}
  K.~P.~Khemchandani, A.~Martínez Torres, A.~Hosaka, H.~Nagahiro, F.~S.~Navarra and M.~Nielsen,
  arXiv:1712.09465 [hep-ph].












\end{thebibliography}
\end{document}